\documentclass{article}
\usepackage{inputenc}
\usepackage[dvips]{graphics}
\usepackage{array,hhline,amssymb,amsthm,color,amsfonts}
\def\L{{\mathfrak{L}}}
\def\O{{\mathcal{O}}}  \def\o{{\mathbf{o}}}
   \def\u{\mathfrak{o}}
     
\def\d12{\raise0.4ex\hbox{\scriptsize 1}\!/\!\lower0.4ex\hbox{\scriptsize 2}}
\def\tr{{\rm tr}} 
\def\buu#1{\buildrel{*}\over{#1}}

\def\bux#1#2{\buildrel{*}\over{#1}_{#2}}

\title{Octonionic Physics}
\date{}
\author{V. Yu. Dorofeev\thanks{Dep. of Math., SPb SUEF,
Sadovaya 21, 191023, St.Petersburg,
Russia, e-mail: friedlab@mail.ru}\\
Laboratory for theoretical physics}
\begin{document}
\maketitle
\begin{abstract}
The physical solutions of Lagrangian of octonionics are researched in the paper. It is shown, the gravitational interaction in Friedmann space and in spherically symmetric space in such model is to be described by pair of charged massless vectorial D-bosons of Minkowski space. It is proposed to use the formalism for the description of jet and supernova.
\end{abstract}
\section{Introduction}
In \cite{Dor1} author proposed the generalization of Lagrangian from Weinberg-Salam theory, devoted to electroweak interactions, on octonionic algebra. This paper is a corollary of the conducted research. Here the physical interpretation of the octonionic Lagrangian from \cite{Dor1} is proposed. The subject of research is the opportunity to describe gravitational interaction on the base of octonionic algebra in the flat Minkowski space.

\section{Algebraic structure of the state space\\ of the extended octonionic space}
The octonionic algebra could be represented by matrices \cite{Dor1}
($i=1,2,3$)
\begin{equation}\label{Zornpr}
\matrix{\Sigma^0=\left(\matrix{\sigma^0&0\cr0&\sigma^0}\right)&\Sigma^i=
\left(\matrix{0&-i\sigma^i\cr i\sigma^i&0}\right)\cr
\Sigma^4=\left(\matrix{-1&0\cr0&1}\right)&
\Sigma^{4+i}=\left(\matrix{0&-\sigma^i\cr -\sigma^i&0}\right)}
\end{equation}
where $\sigma^0$ is a unit matrice and $\sigma^i,i=1,2,3$ is a Pauli matrice:
\begin{equation}\label{matDir}
\sigma^1=\left(\matrix{0&1\cr1&0}\right),\quad
\sigma^2=\left(\matrix{0&-i\cr i&0}\right),\quad
\sigma^3=\left(\matrix{1&0\cr0&-1}\right),\quad
\sigma^0=\left(\matrix{1&0\cr0&1}\right)
\end{equation}

with special multiplication rule
$$\u*\u'=\left(\matrix{\lambda I&A\cr B&\xi I}\right)*
\left(\matrix{\lambda' I&A'\cr B'&\xi' I}\right)=$$
\begin{equation}\label{Dab}
=\left(\matrix{(\lambda\lambda'+\frac12\tr(AB'))I\hfill&\lambda
A'+\xi' A+\frac i2[B,B']\hfill\cr\lambda'B+\xi B'-\frac
i2[A,A']&(\xi\xi'+\frac12\tr(BA'))I\hfill}\right)
\end{equation}

The matrices $A,A',B,B$ with size $(2\times2)$ were introduced, together with unit matrice $I$ of same size and complex numbers $\lambda,\xi,\lambda',\xi'$.

Hermitian conjugation on those matrices is introduced as follows:
\begin{equation}\label{Dab2}
\u=\left(\matrix{\lambda I&A\cr B&\xi I}\right),\qquad
\u^+=\left(\matrix{\lambda^*I&B^+\cr A^+&\xi^*I}\right)
\end{equation}

Multiplication (\ref{Dab}) is defined for a wider matrice class than for matrices from octonionic algebra representation, $\Sigma^a,a=0,1,2,\dots,7$ introduced earlier. Two more matrices could be introduced
\begin{equation}\label{dmo}
f^8=\left(\matrix{0&I\cr
I&0}\right),\quad f^9=\left(\matrix{0& iI\cr-iI&0}\right)
\end{equation}
That maintain multiplication (\ref{Dab}).

So the set of matrices
\begin{equation}\label{lnbaz}
f^a=\Sigma^a,\quad a=0,1,\dots,7,\quad f^8,\quad f^9
\end{equation}
Form a basis of linear space upon the field of complex numbers, which is further called the extended octonionic space, and is denoted for $\overline\O$.

Notice
$$(f^8*f^1)*(f^1*f^8)=0,\quad f^8*(f^1*f^1)*f^8=1$$

Using the multiplication defined in (\ref{Dab}) introduce scalar product in $\overline\O$:
$$(\u_1,\u_2)=\frac12\tr(\u_1^+*\u_2)=\frac12\tr(\left(\matrix{\lambda_1^*I&B_1^+\cr A_1^+&\xi_1^*I}\right)*
\left(\matrix{\lambda_2I&A_2\cr B_2&\xi_2I}\right))$$
\begin{equation}\label{skpro}
=\frac12\tr(\left(\matrix{(\lambda_1^*\lambda_2+\frac12\tr(B_1^+B_2))I\hfill&
\lambda_1^*A_2+\xi_2 B_1^++\frac i2[A_1^+,B_2]\hfill\cr
\lambda_2A_1^++\xi_1^*B_2-\frac
i2[B_1^+,A_2]&(\xi_1^*\xi_2+\frac12\tr(A_1^+A_2))I\hfill}\right))
\end{equation}
$$=\lambda_1^*\lambda_2+\xi_1^*\xi_2+\frac12\tr(B_1^+B_2))
+\frac12\tr(A_1^+A_2)$$

The norm of vector $\o$ in extended octonionic space $\overline\O$
$$\u=\left(\matrix{\lambda^*I&B^+\cr A^+&\xi^*I}\right)$$
is the number
$$||\u||=\sqrt{|\lambda|^2+|\xi|^2+\frac12\tr(A^+A)+\frac12\tr(B^+B))}$$

There is an orthonormal basis with respect to scalar product (\ref{skpro})
\begin{equation}\label{obbaz}
e^0=\left(\matrix{I&0\cr0&0}\right),\quad
e^i=\left(\matrix{0&\sigma^i\cr0&0}\right),\quad
e^4=\left(\matrix{0&0\cr0&I}\right)$$
$$e^{4+i}=\left(\matrix{0&0\cr\sigma^i&0}\right),\quad
e^8=\left(\matrix{0&I\cr0&0}\right),\quad
e^9=\left(\matrix{0&0\cr I&0}\right)
\end{equation}

Obviously, the trace of the production of any three matrices is associative one, i.e.
$$\tr((A*B)*C)=\tr(A*(B*C))=\tr(A*B*C)$$

Introduce a matrice
\begin{equation}\label{mo}
u_0=\frac14\left(\matrix{0&i\sigma^3\cr
0&I}\right)=\frac12(\Sigma^0+\Sigma^4-\Sigma^3-i\Sigma^7)
\end{equation}

Due to high importance of $u_0$ in the construction of octonionic Lagrangian, find the scalar product of this matrice with generatrices $\Sigma^a$ of octonionic algebra 
$$\tr((u_0^+*(\Sigma^a*\Sigma^b)*u_0),\qquad\tr((u_0^+*\Sigma^a)*(\Sigma^b*u_0))$$
(Evidently, the last brackets disposition is the corollary of the present ones and the associativity of trace of a three matrices.)

$$\begin{matrix}{
b=1,\dots,7&\tr(u_0^+*\Sigma^b*u_0)=&-\delta^{3b}\cr
a,b=1,2,3&\tr((u_0^+*\Sigma^a)*(\Sigma^b*u_0))=&\delta^{ab}\cr
a,b=1,2,3&\tr(u_0^+*(\Sigma^a*\Sigma^b)*u_0)=&\delta^{ab}-i\varepsilon^{4ab7}\cr
a,b=5,6,7&\tr((u_0^+*\Sigma^a)*(\Sigma^b*u_0))=
&\delta^{ab}\cr
a,b=5,6,7&\tr(u_0^+*(\Sigma^a*\Sigma^b)*u_0)=
&\delta^{ab}-i\varepsilon^{4ab7}\cr
a=4,b=1,2,3&\tr(u_0^+*(\Sigma^4*\Sigma^b)*u_0)=&0\cr
a=4,b=1,2,3&\tr((u_0^+*\Sigma^4)*(\Sigma^b*u_0))=&0\cr
a=4,b=5,6,7&\tr(u_0^+*(\Sigma^4*\Sigma^b)*u_0)=&-i\delta^{7b}\cr
a=4,b=5,6,7&\tr((u_0^+*\Sigma^4)*(\Sigma^b*u_0))=&-i\delta^{7b}\cr
a=1,2,3,b=5,6,7&\tr((u_0^+*\Sigma^a)*(\Sigma^b*u_0))=&0\cr
a=1,2,3,b=5,6,7&\tr(u_0^+*(\Sigma^a*\Sigma^b)*u_0))=
&-i\varepsilon^{4ab3}\cr
a=5,6,7,b=1,2,3&\tr((u_0^+*\Sigma^a)*(\Sigma^b*u_0))=&0\cr
a=5,6,7,b=1,2,3&\tr(u_0^+*(\Sigma^a*\Sigma^b)*u_0))=
&-i\varepsilon^{4ab3}}
\end{matrix}$$

It is clear from the equations if $a$ and $b$ are equal to 4 the multiplication is associative, otherwise with $a\ne b$ the four-member-production is not associative therefore could be rewritten as follows
\begin{equation}\label{moso}
\tr(u_0^+*\Sigma^a*\Sigma^b*u_0)=\delta^{ab}+ic_c^{ab}A^{ab}_c
\end{equation}

Here, $A^{ab}_c,(c=1,2)$ is antisymmetrical variable equal to $0,\pm1$. In only case when neither $a$ nor $b$ is equal to 4 and special brackets disposition the value of $A^{ab}_c$ differs from zero
\begin{equation}\label{moso2}
a\ne b,\qquad\tr(u_0^+*(\Sigma^a*\Sigma^b)*u_0)=iA^{ab}_2$$
$$\tr((u_0^+*\Sigma^a)*(\Sigma^b*u_0))=iA^{ab}_1=0
\end{equation}
And its non-zero values are
\begin{equation}\label{moso0}
A^{21}_2=A^{65}_2=A^{25}_2=A^{74}_2=A^{74}_1=1
\end{equation}
And if $a=4$ then the order of brackets is does not matter
\begin{equation}\label{moso3}
\tr((u_0^+*\Sigma^4)*(\Sigma^b*u_0))=
\tr(u_0^+*(\Sigma^4*\Sigma^b)*u_0)
\end{equation}
While constants $c_c^{ab}$ are defined by the model of brackets disposition in the production of octonionic operands, i.e. by relief of non-associativity. Three models of it are proposed below.

{\bf Probability model of non-associativity relieving.}
\begin{equation}\label{vm0}
\tr(A*B*C*D)=p_1\tr(((A*B)*C)*D)+p_2\tr(A*(B*C)*D)
\end{equation}
$$p_1+p_2=1$$

Here the frequencies $p_1$ and $p_2$ are introduced. Their values depend on the number of brackets permutations leading to the same result in (\ref{vm0}). For example, in (\ref{vm0}) it is natural to admit $c_1^{ab}=p_1=3/4,c_2^{ab}p_2=1/4$ because three different types of brackets disposition lead to the same result in the first member in the right part of the equation (14) whereas only one type - in the second member of the right part of (\ref{vm0}). 

Similarly, the probability model of non-associativity relieving for (\ref{moso} gives
\begin{equation}\label{vm01}
\tr(u_0^+*\Sigma^a*\Sigma^b*u_0)=\delta^{ab}+\frac{3i}4A^{ab}_1+\frac i4A^{ab}_2
\end{equation}

The definition given in (\ref{vm0}) is also applicable for greater number of elements.

{\bf Minimal model of non-associativity relieving.}
\begin{equation}\label{vm2}
f(ABCD)\tr(A*B*C*D)=\min_{()}(f(ABCD)\tr(A*B*C*D))
\end{equation}

The right part of (\ref{vm2}) means the choice of brackets disposition is done to minimize $f(ABCD)\tr(A*B*C*D)$. There the coefficient $f(ABCD)$ (which could be equal to 1) is introduced before $\tr$. Notice, the coefficient before $\tr$ does not necessarily require its incorporation in the definition of the minimal model.

{\bf The maximal model} is to be defined similarly.

Obviously, in associative case the same result is obtained to that of the models defined above.

It is easy to ensure, with $i\ne j,k\ne l$
\begin{equation}\label{chp5}
\frac14\tr(\Sigma^i*\Sigma^j*\Sigma^k*\Sigma^l)=
-\varepsilon^{ijkl}\delta^{ik}\delta^{jl}+A^{ijkl}
\end{equation}
where $A^{ijkl}$ equals to $\pm1,0$ regarding to the brackets disposition in production and the values of indices.

\section{Lagrangian of octonionics}
In \cite{Dor1} the generalization of Weinberg-Salam Lagrangian on octonionic algebra is proposed as follows: ($a,b=0,1,\dots,7,k=1,2,\dots,7$):
$$\L_{oct.}=\L_f+(\partial_\mu\bux\Psi\varphi-
\frac i2q^aA_\mu^{a}\bux\Psi\varphi*\Sigma^{a})
*(\partial^\mu\Psi_\varphi+
\frac i2q^{b}A^{\mu(b)} \Sigma^{b}*\Psi_\varphi)$$
$$+\frac i2\overline L*\gamma_\mu(\overrightarrow\partial^\mu
L+\frac i2c_Lq^kA^{\mu(k)}\Sigma^k*L+\frac
i2c_{L_0}q^0A^{\mu(0)}L)$$
$$-\frac i2\overline L*\gamma_\mu(\overleftarrow\partial^\mu L-
\frac i2c_Lq^kA^{\mu(k)}\Sigma^{k}*L-\frac
i2c_{L_0}q^0A^{\mu(0)}L)$$
$$+\frac i2\overline R\gamma_\mu(\overrightarrow\partial^\mu R+
iq^0A^{\mu0}R)-\frac i2\overline
R\gamma_\mu(\overleftarrow\partial^\mu R-iq^0A^{\mu0}R)$$
\begin{equation}\label{vst}
-\tilde h\overline L*\Psi_\varphi R-\tilde h\overline
R\bux\Psi\varphi*L)+m^2||\Psi_\varphi||^2-\frac f4||\Psi_\varphi||^4
\end{equation}

There following denotes are done. Free fields Lagrangian $\L_f$ \begin{equation}\label{svs0}
\L_f=-\frac14F^0_{\mu\nu}F^{\mu\nu(0)}-\frac14\tr(F_{\mu\nu}^k*F^{\mu\nu(k)})
$$ $$+\frac1{16}f^{ijkl}(A_\mu^iA_\nu^j-A_\nu^j
A_\mu^i)(A^{\mu(k)}A^{\nu(l)}-A^{\nu(k)}A^{\mu(l)})
\end{equation}
\begin{equation}\label{l04}
f^{ijkl}=q^{ij}q^{kl}\tr(\Sigma^i*\Sigma^j*\Sigma^k*\Sigma^l)
\end{equation}
$$F^0_{\mu\nu}=\partial_\mu A^0_\nu-\partial_\nu A^0_\mu,\qquad
F_{\mu\nu}^k=\partial_\mu A_\nu^k-\partial_\nu A_\mu^k-\varepsilon^{ijk}q^{ij}(A_\mu^iA_\nu^j-A_\nu^iA_\mu^j)$$

The value $f^{ijkl},i,j,k,l=1,\dots,7$ reflects the non-associate character of free octonionic fields Lagrangian.

Left and right spinor components
\begin{equation}\label{lpm}
\frac12(1+\gamma^5)\Psi=L,\qquad\frac12(1-\gamma^5)\Psi=R,\qquad\gamma^5=i\gamma^0\gamma^1\gamma^2\gamma^3
\end{equation}

$\Psi,\Psi_\varphi$ are the vectors from the generalized octonionic state space, $q^a,c_L,c_{L_0}$ are some numbers, caused by normalization; $\gamma^\mu$ -- are Dirac matrices ($\mu=0,1,2,3$):
\begin{equation}\label{matDir2}
\gamma^0=\left(\matrix{I&0\cr0&-I}\right),\qquad
\gamma^i=\left(\matrix{0&\sigma^i\cr-\sigma^i&0}\right), \quad
i=1,2,3
\end{equation}

The summing up is carried out with metric tensor of Minkowski space $\eta_{\mu\nu}$ with signature $(1,-1,-1,-1)$, by different-case Greek indices, while the simple summing up is conducted by the same-case indices.

It is also shown in \cite{Dor1} that the vectors from extended state space $\bar\O$
\begin{equation}\label{matDir3}
\Psi_\varphi=\Psi_0=\frac{m}{\sqrt 2f}\left(\matrix{0&i\sigma^3\cr
0&I}\right)=\frac{\sqrt2m}{\sqrt f}u_0,\qquad R(x)=e^-_R(x)$$
$$\Psi=\left(\matrix{(\alpha_1\nu+\alpha_2e)I&A_1\nu(x)+A_2e(x)\cr
B_1\nu(x)+B_2e(x)&(\beta_1\nu+\beta_2e)I}\right)
\end{equation}
and numbers
$$c_{L_0}=-1,\quad c_0^2=\frac{32}{257},\quad
c_L=(c_0^2)^{-1},\tilde h=2\sqrt2h/c_0,$$
$$q^1=q^2=q^3=q^{12}=g,\qquad q^0=g'$$
With $k=1,2,3$ in (\ref{lpm}) give model from electron-neutrino sector of Weinberg-Salam theory with group charge of $SU(2)$-symmetry $g$ and charge $g'$ of symmetry $U(1)$, Higgs field $\Psi_0$, and the constant of interaction of Higgs field with matter fields $h$, see \cite{Okun}.

Once more referring to \cite{Dor1}, the expression for octonionic Lagrangian in case $\Psi_\varphi=\Psi_0$ is obtained there
$$\L_{oct}=\L_f+
\frac{q^{(k)2}m^2}{2f}A_\mu^{k}A^{\mu(k)}+\frac{o^{ij}m^2}{2f}A_\mu^{i}A^{\mu(j)}+
\frac{g^{(1)2}m^2}{2f}B_\mu B^\mu
-\frac{gg^{(1)}m^2}fA_\mu^3B^\mu$$
$$+\frac{g^{(1)}}2\overline\nu_L\gamma_\mu B^\mu\nu_L+
\frac{g^{(1)}}2\overline e_L\gamma_\mu B^\mu e_L+\frac g2\overline
e_L\gamma_\mu A^{\mu3}e_L-\frac g2\overline\nu_L\gamma_\mu
A^{\mu3}\nu_L$$
$$-\frac g2\overline\nu_L\gamma_\mu e_L(A^{\mu1}-iA^{\mu2})-
\frac g2\overline e_L\gamma_\mu\nu_L(A^{\mu1}+iA^{\mu2})$$
$$+\frac i2(\overline e_L\gamma_\mu\partial^\mu e_L-\partial^\mu\overline
e_L\gamma_\mu e_L)+\frac
i2(\overline\nu_L\gamma_\mu\partial^\mu\nu_L-
\partial^\mu\overline\nu_L\gamma_\mu\nu_L)+\frac{m^4}f$$
$$+\frac i2(\overline e_R\gamma_\mu\partial^\mu
e_R-\partial^\mu\overline e_R\gamma_\mu e_R)+g^{(1)}\overline
e_R\gamma_\mu B^\mu e_R-\frac{\sqrt2hm}{\sqrt f}(\overline
e_Le_R+\overline e_Re_L)$$
$$-q^4A^{\mu(4)}
(\kappa_1\overline\nu_L\gamma_\mu\nu_L- \kappa_2\overline
e_L\gamma_\mu e_L)-
\frac32q^6A^{\mu(6)}\overline e_L\gamma_\mu e_L$$
\begin{equation}\label{plna}
-\frac54(q^6A^{\mu(6)}+iq^5A^{\mu(5)})\overline\nu_L\gamma_\mu
e_L-\frac54(q^6A^{\mu(6)}-iq^5A^{\mu(5)})\overline e_L\gamma_\mu
\nu_L
\end{equation}
Where the following notation is used $ic_2^{ij}A^{ij}q^iq^j=o^{ij}$. 

So, the final Lagrangian contains non-associative elements along with associative.

\section{Octonionic Lagrangian research}
1. Consider non-associative summands from Lagrangian (24). First of all, it is the quadratic member by the fields $A_\mu^k$ 
\begin{equation}\label{ina1}
\frac{o^{ij}m^2}{2f}A_\mu^{i}A^{\mu(j)}=ic_k^{ij}A^{ij}_k \frac{q^iq^jm^2}{2f}A_\mu^{i}A^{\mu(j)}
\end{equation}

Apply probability model of non-associativity relieving to this member. Assume for non-zero components $c^{ij}_1=c^{ji}_1=3/4, c^{ij}_2=c^{ji}_2=1/4$ Hence due to symmetry of the expression $q^iq^jA_\mu^{i}A^{\mu(j)}$ with respect to $i,j$ (there is no summing up by them) and the anti-symmetry of multiplier $A^{ij}_k$ (again with respect to $i,j$) this member equals zero.

2. The free fields Lagrangian Lf contains non-associativity of fourth order with respect to fields $A_\mu$.
\begin{equation}\label{ina2}
f^{ijkl}(A_\mu^iA_\nu^j-A_\nu^j
A_\mu^i)(A^{\mu(k)}A^{\nu(l)}-A^{\nu(k)}A^{\mu(l)})
\end{equation}

Formally the summand in Lagrangian is to be considered as potential energy (similarly to $\lambda\varphi^4$), therefore the maximal scheme of brackets disposition is applicable here (assuming $L=T-V$). When considering that member with minus then the model of brackets would be minimal, therefore implying the physical application of the problem, in what follows the brackets disposition according to $\min$ or $\max$ rule is called {\bf potential model}.

3. Lagrangian in (\ref{plna}) contains a neutral current
\begin{equation}\label{is3}
-q^4A^{\mu(4)}(\kappa_1\overline\nu_L\gamma_\mu\nu_L- \kappa_2\overline
e_L\gamma_\mu e_L)
\end{equation}
which interacts with left spinors only. In this manner the neutral vectorial boson $C_\mu$ is defined
\begin{equation}\label{is2}
q_CC_\mu=-q^4A^{\mu(4)}
\end{equation}

4. Lagrangian (\ref{plna}) contain current
\begin{equation}\label{is1}
-\frac54\overline\nu_L\gamma_\mu(q^6A^{\mu(6)}+iq^5A^{\mu(5)})
e_L-\frac54(q^6A^{\mu(6)}-iq^5A^{\mu(5)})\overline e_L\gamma_\mu\nu_L
\end{equation}
which indicates the necessity of introduction of two oppositely charged vectorial bosons
\begin{equation}\label{is20}
q_DD_\mu=-\frac54q^6A^{\mu(6)}-i\frac54q^5A^{\mu(5)}$$
$$q_D\buu D_\mu=-\frac54q^6A^{\mu(6)}+i\frac54q^5A^{\mu(5)}
\end{equation}
With charge $\pm q_D$. 

Hence, it is necessary to modify the current $q^6A^{\mu(6)}\overline e_L\gamma_\mu e_L$ in as follows
\begin{equation}\label{is30}
-\frac32q^6A^{\mu(6)}\overline e_L\gamma_\mu e_L=$$
$$-\frac34\overline e_L\gamma_\mu(q^6A^{\mu(6)}+iq^5A^{\mu(5)})e_L) -\frac34(q^6A^{\mu(6)}-iq^5A^{\mu(5)})\overline e_L)\gamma_\mu e_L
\end{equation}

5. Lagrangian (\ref{plna}) contains Lagrangians of left electron fields, which interact with vector field $A^{\mu(6)}$
\begin{equation}\label{is4}
\frac i2(\overline e_L\gamma_\mu\partial^\mu e_L-\partial^\mu\overline e_L\gamma_\mu e_L)-\frac32q^6A^{\mu(6)}\overline e_L\gamma_\mu e_L
\end{equation}
which is equivalent (in analogue to electro-magnetic field) to the expression with long derivative
\begin{equation}\label{is5}
\partial^\mu-i\frac34q^6A^{\mu(6)}
\end{equation}
And given Remark 4 the long derivative transforms to

\begin{equation}\label{is6}
\partial^\mu-i\frac34q^6A^{\mu(6)}-\frac34q^5A^{\mu(5)}
\end{equation}

To summarize, the Lagrangian of matter fields, which interact with vector fields can be rewritten as follows
\begin{equation}\label{is7}
\frac i2\overline e_L\gamma_\mu\partial^\mu e_L-\frac i2\partial^\mu\overline e_L\gamma_\mu e_L-\frac32q^6A^{\mu(6)}\overline e_L\gamma_\mu e_L=$$
$$=\frac i2\overline e_L\gamma_\mu((\partial^\mu-i\frac34q^6A^{\mu(6)}-\frac34q^5A^{\mu(5)}) e_L)-$$
$$-\frac i2((\partial^\mu+i\frac34q^6A^{\mu(6)}-\frac34q^5A^{\mu(5)})\overline e_L)\gamma_\mu e_L
\end{equation}

\section{Tetrad representation}
Consider a differentiable Riemannian manifold $M$ with local coordinates $x^\mu(p)$, $p\in M,\mu=0,1,2,3$ and metrics
\begin{equation}\label{mpr}
ds^2=g_{\mu\nu}dx^\mu dx^\nu
\end{equation}
Covariant derivative is defined as follows
\begin{equation}\label{kp}
A^\mu_{;\nu}=A^\mu_{,\nu}+\Gamma^\mu_{\nu\lambda}A^\lambda
\end{equation}
And it is a vector. Christoffel symbols of second sort
\begin{equation}\label{sk}
\Gamma^\mu_{\nu\lambda}=\frac12g^{\mu\kappa}(g_{\mu\kappa,\nu}+
g_{\nu\kappa,\lambda}-g_{\lambda\nu,\kappa})
\end{equation}
Do not complete a tensor. However, what is subjected to tensors transformation law is the Christoffel symbols combination - Riemannian tesnor, which is defined according to \cite{Eisenhart}
\begin{equation}\label{tkr}
R^\tau_{\mu\nu\lambda}=\Gamma^\tau_{\mu\lambda,\nu}-
\Gamma^\tau_{\mu\nu,\lambda}+\Gamma^\tau_{\sigma\nu}\Gamma^\sigma_{\mu\lambda}-
\Gamma^\tau_{\sigma\lambda}\Gamma^\sigma_{\mu\nu}
\end{equation}

Quadratic form (\ref{mpr}) could be diagonalized in some new coordinate system. Associating surface parameters with physical space-time parameters, and assuming in some coordinates metrics is transformed to the Minkowski space metric type:
\begin{equation}\label{mpm}
ds^2=c^2dt^2-dx^2-dy^2-dz^2=\eta_{ab}dx^adx^b
\end{equation}

In each point of space-time introduce the tetrads $h^\mu_a=\partial x^\mu/\partial x^a,a=0,1,2,3$ which implement such transformation. Tetrads comply with a natural orthogonality property 
\begin{equation}\label{v010}
h^b_\mu h^\mu_{a,\nu}=\delta^b_a,\qquad h^{\mu(a)}h^\nu_a=g^{\mu\nu} \end{equation}
and with the properties to lift and to sink indices.
\begin{equation}\label{v12}
\delta A^\mu=\delta(A^ah^\mu_a)=\delta A^ah^\mu_a+
A^ah^\mu_{a,\nu}\delta x^\nu=\Gamma^\mu_{\nu\lambda}A^\nu\delta x^\lambda \end{equation}

In each point of Riemannian space contravariant vector A... could be considered in diagonal coordinates system $A^a=A^\mu
h^a_\mu$. ”читыва€ $h^b_\mu h^\mu_{a,\nu}+h^b_{\mu,\nu} h^\mu_a=0$, find
\begin{equation}\label{t2}
\delta A^b=\gamma^b_{ac}A^a\delta x^c,\qquad\gamma^b_{ac}=\eta^{bd}
\gamma_{dac}=h_{\mu(d);\nu}h^\mu_ah^\nu_c
\end{equation}
where $\gamma^b_{ac}$ are Ricci coefficients.

Well-known spinor transformation $\Psi(x)$ in case of one likes Lorentz transform \cite{Ahieser}
\begin{equation}\label{v012}
\Psi\to\Psi+\frac12\omega^{ab}\sigma^{ab}\Psi,\qquad\sigma^{ab}=\frac14[\gamma^a,\gamma^b],\qquad a,b=0,1,2,3
\end{equation}
Where the constant matrix of infinitesimal transformation $\omega^{ab}$, defining Lorentz rotation, is introduced. Dirac matrices comply with the following multiplication rule on matrices $\sigma^{ab}$:
\begin{equation}\label{v012d}
\gamma^a\sigma^{bc}=\frac14\gamma^a[\gamma^b,\gamma^c]=\frac12\eta^{ab}\gamma^a-\frac12\eta^{ac}\gamma^b-\frac i2\varepsilon^{dabc}\gamma^5\gamma_d
\end{equation}

It is well-known, $\overline\Psi\gamma^\mu\Psi$ is a vector regarding to overall-coordinates transformation \cite{Ahieser}. The latter imposes restrictions on spinors transformations. To concord overall-coordinate transformation with Lorentz transform replace the infinitesimal matrix $\omega^{ab}$ with the matrix depending on $x,\omega^{ab}=\omega^{ab}(x)$ in analogue with local gauge symmetry \cite{Utiyama}. As a result, find the general expression for derivative in locally flat Minkowski space, which would be a vector to overall-coordinates transformation
\begin{equation}\label{ivsok1}
\partial_a\Psi\to D_a\Psi=\partial\Psi/\partial x^a-i\Phi_a\psi-\Gamma_a\Psi
\end{equation}

This transformation first appears in the work of Fock-Ivanenko \cite{FokIvanenko}. It was shown there, that $\Gamma_a$ is a real value depending on rotation coefficients $\gamma_{abc}$ and Dirac matrices and $\Phi_a$ is a real value proportional to a unit matrix. Furthermore in \cite{FokIvanenko} $\Phi_a$ is identified exceptionally with electromagnetic field. Matrix $\Gamma_a$\begin{equation}\label{v013}
\Gamma_a=-\frac12\gamma_{abc}\sigma_{bc}
\end{equation}

Particularly, considered in orthogonal coordinates, gravitational field in Riemannian space with metrics
\begin{equation}\label{ivsok2}
ds^2=H^{(0)2}dx^{(0)2}-
H^{(1)2}dx^{(1)2}-H^{(2)2}dx^{(2)2}-H^{(3)2}dx^{(3)2}
\end{equation}
And given (\ref{v012d}) lead to the following
\begin{equation}\label{ivsok20}
-\gamma^a\Gamma_a=
\frac14h^\mu_ah^\nu_bh_{(c)\nu;\mu}(\eta^{ab}\gamma^a-\eta^{ac}\gamma^b- i\varepsilon^{dabc}\gamma^5\gamma_d)\end{equation}

Recall to the metrics (\ref{ivsok2}) is diagonal therefore tetrads $h_{\mu(a)}$ are also diagonal, then with different $a,b,c$ it takes
\begin{equation}\label{ivsok201}
\Gamma^\lambda_{\mu\nu}h_{c\lambda}h^\mu_bh^\nu_c=\frac12g^{\mu\kappa}(g_{\mu\kappa,\nu}+
g_{\nu\kappa,\lambda}-g_{\lambda\nu,\kappa})h_{c\lambda}h^\mu_bh^\nu_c=0
\end{equation}
hence
\begin{equation}\label{ivsok202}
-\gamma^a\Gamma_a=\frac14h^\mu_ah^{\nu(a)}h_{(c)\nu;\mu}\gamma^c-\frac14
h^{\mu(a)}h^{\nu(b)}h_{(a)\nu;\mu}\gamma^b$$
$$=\frac14h^\mu_{c;\mu}\gamma^c+\frac14h^{\mu(a)}h^\nu_{b;\mu}h_{(a)\nu;\mu}\gamma^b=\frac12h^\mu_{c;\mu}\gamma^c
\end{equation}
and Dirac equation is converted to Ivanenko-Sokolov equation \cite{SokolovIvanenko}:
\begin{equation}\label{ivsok3}
(i\gamma^a(H^a)^{-1}(\partial_a-i\Phi_a+\frac12
\partial_a\left(\ln\frac{\sqrt{-g}}{H^a}\right))+m)\psi=0
\end{equation}
Assume Dirac equation is rewritten in Cartesian coordinate system. In that case we identify field $\Phi_a$ с $A^6_a$ with $A^6_a$ and field $\Gamma_a$ with $A^5_a$ (in absence of EMF). This analogy follows from Dirac equations (\ref{is7}) and (\ref{ivsok3}) comparison

So, in orthogonal coordinates it takes (\ref{ivsok2}):
\begin{equation}\label{pdg}
q_DA_a^5(x)=\frac12\partial_a\left(\ln\frac{\sqrt{-g}}{H^a}\right),
\qquad \sqrt{-g}=H_0H_1H_2H_3
\end{equation}

Therefore
\begin{equation}\label{a52}
H^a=e^{\frac23q_D\oint A_b^5dx^b-2\int  A_a^5dx^a},\qquad
\sqrt{-g}=e^{\frac23q_D\oint A_b^5dx^b}
\end{equation}

(In the integral of kind $\int$ it is assumed the absence of summing up by index $a$, and in the integral of kind $\oint$ vice versa is assumed the absence of summing up by $b$.)

Actually curvilinear space appears as a solution method for the Dirac equation in external field $A_\mu^5$, but as free variables $x^k$ have clear physical meaning, it is so proved the generalization of Weinberg-Salam theory on Caley octaves algebra, by means proposed in this paper, is equal to introduction of curvilinear space which is the background for matter fields consideration.

\section{Friedmannian space}
In this section it is shown in the flat Friedmannian space there is a self-consistent solution of octonionics Lagrangian.

Consider homogeneous and isotropic Universe

\begin{equation}\label{oiv}
ds^2=dx^{(0)2}-a^2(t)(dx^{(1)2}+dx^{(2)2}+dx^{(3)2})=
a^2(\eta)(d\eta^2-dl^2),
\end{equation}
with the conformal time $dt=a(\eta)d\eta$, and wit
\begin{equation}\label{f1}
g_{00}=a^2(\eta),\quad g_{\alpha\beta}=a^2(\eta)\eta_{\alpha\beta}
\end{equation}

By means of (\ref{sk}) find all the nonzero components of Christoffel symbol:
\begin{equation}\label{f2}
\Gamma^0_{00}=\frac{a'}{a^3},\quad\Gamma^0_{\alpha\beta}=-\frac{a'}{a^3}g_{\alpha\beta},\quad\Gamma^\alpha_{0\beta}=\frac{a'}a\delta^\alpha_\beta
\end{equation}
and estimate the value 
\begin{equation}\label{f3}
G=g^{\mu\nu}(\Gamma^\lambda_{\mu\nu}\Gamma^\kappa_{\lambda\kappa}- \Gamma^\lambda_{\mu\kappa}\Gamma^\kappa_{\nu\lambda})
=\frac{6{a'}^2}{a^4},
\end{equation}
where the stroke means the derivative with respect to conformal time.

From (\ref{pdg}) it takes 
\begin{equation}\label{f30}
q_DA_\mu^5=(\frac{da}{a^2d\eta},\vec 0)
\end{equation}

Given $A_\mu^5$ and $A_\mu^6$ are united in one vectorial field $D_\mu$, deduce $A_\mu^5=A_\mu^6$ in case under consideration.

Assume 
\begin{equation}\label{f4}
\frac{q^{47}q^{56}}{q_D^2}A^{(4)\mu}A^{(7)}_\mu=Constant=-\frac32\gamma
\end{equation}
then non-associative on fields Lagrangian member (\ref{plna}) can rewritten
\begin{equation}\label{f6}
\frac1{16}f^{ijkl}(A_\mu^iA_\nu^j-A_\nu^j
A_\mu^i)(A^{\mu(k)}A^{\nu(l)}-A^{\nu(k)}A^{\mu(l)})$$
$$=4q^{47}q^{56}A^{(4)\mu}A^{(7)}_\mu A^{(5)\nu}A^{(6)}_\nu=
-6\gamma\left(\frac{da}{a^2d\eta}\right)^2=-6\gamma\left(\frac{a'}{a^2}\right)^2
\end{equation}
Estimated using potential scheme of non-associativity relieving. Associate this expression with gravitation field Lagrangian\begin{equation}\label{f7}
\L_{grav.}=4q^{47}q^{56}A^{(6)\mu}A^{(6)}_\mu A^{(5)\nu}A^{(5)}_\nu=
-\gamma G
\end{equation}
While the gravitation field action is
\begin{equation}\label{f8}
S=\int d^4x(-\gamma G+\L_{mat.})=\int d^4x(-\gamma G+\L_{mat.})-\gamma\int d^4xdiv\Gamma$$
$$=\int d^4x(-\gamma R+\L_{mat.})
\end{equation}
Where $R$ is the curvature. Here the divergent members are introduced, which complete Lagrangian $G$ to curvature, the surface member is assumed to be zero. As a corollary Einstein equation is obtained
\begin{equation}\label{uren}
R_{\mu\nu}=\frac1\gamma T_{\mu\nu}
\end{equation} 

However, the question of legitimacy of (\ref{f4}) and (\ref{f30}) is still left untouched, that is why further research of the free octonionic Lagrangian L is needed. Because only the long-range part of it, without EMF, is important for the research, so eliminate $A_\mu^k,k=01,2,3$. Let $q^5A^5_\mu=q^6A^6_\mu=q_DD_\mu$ then
\begin{equation}\label{uel1}
F_{\mu\nu}^D=D_{\nu,\mu}-D_{\mu,\nu}
\end{equation}

And Euler-Lagrange equation performs
\begin{equation}\label{uel10}
F^{\mu\nu(D)}_{,\mu}+
(m^2_D+q^{47}q^{56}A^{\mu(4)}A^{\nu(7)})D^\nu=0
\end{equation}

Assume mass $m_D$ to be that abolishing the expression in brackets (\ref{uel10})
\begin{equation}\label{uel1d}
m^2_D=-q^{47}q^{56}A^{\mu(4)}A^{\nu(7)})=\frac32q^2_D\gamma
\end{equation}

Once expression in brackets equals zero find the solution for vectorial $D$-boson
\begin{equation}\label{uel2}
F^{\mu\nu(5)}_{,\mu}=F^{\mu\nu(6)}_{,\mu}=0,\qquad A^5_\mu=A^6_\mu=(g(t),\vec0)
\end{equation}
where $g(t)$ could be any time-dependent function. So $D$-bosons are proved to be massless vectorial charged particles.

Consider equation for $C$- and $E$-bosons (denote $\dot a=\frac{da}{dt}$)
\begin{equation}\label{uel3}
F^{\mu\nu(C)}_{,\mu}+m_C^2C^\nu+q^{47}q^{56}\frac{\dot a^2}{a^2}E^\nu=0$$
$$F^{\mu\nu(E)}_{,\mu}+m_E^2E^\nu+q^{47}q^{56}\frac{\dot a^2}{a^2}C^\nu=0
\end{equation}

Assuming $C$ and $E$ - boson masses so that
\begin{equation}\label{uel40}
\matrix{\displaystyle m_C^2C^\nu+ q^{47}q^{56}\frac{\dot a^2}{a^2}E^\nu\approx m_C^2C^\nu\cr\cr\displaystyle
m_E^2E^\nu+q^{47}q^{56}\frac{\dot a^2}{a^2}C^\nu\approx m_E^2E^\nu}
\end{equation}
come to the expression for free vectorial massive particles. For solutions
\begin{equation}\label{uel5}
C_\mu=e^{ikx}c_\mu,\qquad E_\mu=e^{-ikx}e_\mu
\end{equation}
with constant polarization vectors $c_\mu$ and $e_\mu$ obtain
\begin{equation}\label{uel50}
C_\mu E_\mu=c_\mu e_\mu=Constant
\end{equation}
come to condition (\ref{f4}).

From octonionics Lagrangian it is inferred $E_\mu$-bosons decompose to particles and anti-particles almost immediately (assumed their mass is large), however, the proposed model deals with the pair $C\bar E$ - a particle and an anti-particle in the bound state. Moreover the given state defines vacuum, this is why it is assumed the decomposition is denied in ordinary terms. Presumably, special terms are early cosmology and closely to black holes particles birthes.

For the initially posed question was the Friedmann type solutions existence on the octonionics Lagrangian, the problem could be deemed as solved.

\section{Schwarzchild solution}
Consider the space where there is a massive spherically symmetrical object. Let this object be the source of octonionic field. On great distances from the object the electroweak interaction could be eliminated therefore the object could be merely the source of the senior octonionic fields. Due to symmetry of the problem it is acceptable if on the great distances there are octonionic fields, caused by the object, then they are produced by vector-potential $A_\mu^k=A_\mu^k(r),k=4,5,6,7$.

Let, e.g. the electron, moving in the space. The space where the electron is moving is by definition the Minkowski space. In spherically symmetrical coordinates the metrics in the space forms
\begin{equation}\label{Mprs}
ds^2=dt^2-dr^2-r^2(\sin^2\theta d\varphi^2+d\theta^2)
\end{equation}

Restrict the consideration by left spinors. Rewrite its the equation of moving in massive source octonionic field, assumed it does not interact with the fields $A^{4,7}_\mu$
\begin{equation}\label{ivsok31}
(i\gamma^0\partial_0+i\gamma^r(\partial_r-\vec\Sigma\cdot\hat{\vec L}-\frac{3i}4q^6A^6_r-\frac34q^5A^5_r)+m)\psi=0
\end{equation}
here denote
$$\gamma^r=\gamma^1\sin\theta\cos\varphi+
\gamma^2\sin\theta\sin\varphi+\gamma^3\cos\theta,\quad
\vec\Sigma=\left(\matrix{\vec\sigma&0\cr0&\vec\sigma}\right)$$
where $\vec L=\vec r\times\vec p$ is the angular moment operator. \cite{Wheeler}.

In solutions (\ref{ivsok31}) the spherically-symmetrical part is selected, that responses to the proper value of angular momentum, hence comparison of (\ref{ivsok31}) with Ivanenko-Sokolov equation (\ref{ivsok3}) gives the idea that formally its solution could be found in the curvilinear space-time
\begin{equation}\label{Shwm00}
ds^2=H_0^2(r)dt^2-H^2_1(r)dr^2-r^2(\sin^2\theta d\varphi^2+d\theta^2)
\end{equation}
(denoted $\bar g=-H_0^2H_1^2$)
\begin{equation}\label{ivsok32}
-\frac34q^5A^5_r=\frac12\partial_a\left(\ln\frac{\sqrt{-\bar g}}{H^a}\right))=\frac{H_{0,r}}{2H_0},\qquad A^5=(0,A^5_r,0,0)
\end{equation}

For a weak field it is well-known 
\begin{equation}\label{slp}
g_{00}=1-r_g/r=f^2
\end{equation}
So assume $H_0$ to be known and find $H_1(r)$. Approximately, with great $r$, assume $H_1=1+C/r^n,n>1$. Find gravitational field Lagrangian $G_{gr}$ in metrics (\ref{Shwm00})
\begin{equation}\label{gshv10}
G_{gr}=g^{\mu\nu}(\Gamma^\lambda_{\mu\nu}\Gamma^\kappa_{\lambda\kappa}- \Gamma^\lambda_{\mu\kappa}\Gamma^\kappa_{\nu\lambda})
=\frac2{r^2H_1^2}+\frac{4H_{0,r}}{rH_1^2H_0}-\frac{H_{0,r}H_{1,r}}{H_1^3H_0}
\end{equation}

On the other hand according to the octonionic theory the gravitational field Lagrangian is defined as
\begin{equation}\label{gshv3}
-\gamma \L_{gr}=G_{pl}+C_0q^{(5)2}A_\mu^5A^{(5)\mu}= \frac2{r^2}+C_0\frac{4H_{0,r}^2}{9H_0^2}
\end{equation}
(Here the member with zero-curvature, $G_{pl}$, is added, which responses to the zero curvature of the value $G$ in curvilinear coordinates and the constant $C_0$ responds to the convolution $A^{(4)\mu}A^7_\mu$.) 

To obtain the same infinitesimal order of (\ref{gshv10}) and (\ref{gshv3}) when $r$ it is required (accurate within next infinitesimal orders in decomposition by $r$ powers)
\begin{equation}\label{gshv4}
H_0H_1=1,\qquad g_{11}=(1-r_g/r)^{-1}
\end{equation}

Finally the spherically-symmetrical Schwarzchild metrics is obtained
\begin{equation}\label{Shwm0}
ds^2=(1-\frac{r_g}r)dt^2-\frac{dr^2}{1-\frac{r_g}r}-
r^2(\sin^2\theta d\varphi^2+d\theta^2)
\end{equation}
where $r_g=2M/r$ is gravitational radius. Then from the formal, built up in the paper, point of view, it is necessary to consider complementary vectorial field $D_\mu$ so that
\begin{equation}\label{DDm}
q_DD_\mu=q_6A_\mu^6+iq_5A_\mu^5,\qquad q_5A_\mu^5=q_6A_\mu^6=f_{,r}/f
\approx \frac{r_g}{2r^2}
\end{equation}

However, the question of admissibility of (\ref{DDm}) arose therefore research the free Lagrangian $\L_f$. Once only its long-range piece without $A_\mu^k,k=0,1,2,3$ is under interest, then eliminate $q^5A^5_\mu=q^6A^6_\mu=q_DD_\mu$ and it takes
\begin{equation}\label{uel12}
F_{\mu\nu}^D=D_{\nu,\mu}-D_{\mu,\nu}
\end{equation}

The Euler-Lagrange equations
\begin{equation}\label{uel13}
F^{\mu\nu(D)}_{,\mu}+
(m^2_D+q^{47}q^{56}A^{\mu(4)}A^{\nu(7)})D^\nu=0
\end{equation}

Let $m_D$ mass is that abolishing the expression in brackets in (\ref{uel13})
\begin{equation}\label{uel1d5}
m^2_D+q^{47}q^{56}A^{\mu(4)}A^{\nu(7)}=0
\end{equation}

Once the expression in brackets equals zero find the solution for vctorial $D$-boson

\begin{equation}\label{uel20}
F^{\mu\nu(D)}_{,\mu}=0,\qquad A^5_\mu=A^6_\mu=(0,g(r),0,0)
\end{equation}
where $g(r)$ is any coordinates-dependent function. So $D$-bosons are proved to be massless vectorial charged particles.

Consider equations for $C$- and $E$-bosons\begin{equation}\label{uel30}
F^{\mu\nu(C)}_{,\mu}+m_C^2C^\nu+q^{47}q^{56}\frac{r_g^2}{4r^4}E^\nu=0$$
$$F^{\mu\nu(E)}_{,\mu}+m_E^2E^\nu+q^{47}q^{56}\frac{r_g^2}{4r^4}C^\nu=0
\end{equation}

Assume $C$- and $E$-bosons' masses are so that
\begin{equation}\label{uel4}
m_C^2C^\nu+ q^{47}q^{56}\frac{r_g^2}{4r^4}E^\nu\approx m_C^2C^\nu,\qquad m_E^2E^\nu+q^{47}q^{56}\frac{r_g^2}{4r^4}C^\nu\approx m_E^2E^\nu
\end{equation}
then come to the free vectorial massive particles equation. For solutions
\begin{equation}\label{uel501}
C_\mu=e^{ikx}c_\mu,\qquad E^\mu=e^{-ikx}e^\mu
\end{equation}
With constant polarization vectors $c_\mu$ and $e_\mu$ it takes
\begin{equation}\label{uel57}
C_\mu E^\mu=c_\mu e^\mu=Constant
\end{equation}
and condition (\ref{f4}) is obtained.

As it was earlier assumed, the pairs $C\bar E$ in ordinal conditions comprise a stable state.

For the initially posed question was the Friedmann type solutions existence on the octonionics Lagrangian, the problem could be deemed as solved.

Formally, in the paper it is shown the existence of the approximate octonionic field, which could be appropriate one for physical reality. These considerations are deemed by author as deserving the most scrupulous attention and further analysis of octonionic gravity approach, proposed above.
\section{Jet phenomenology}
In astronomy the beautiful phenomenon is well-known, when from the center of a galaxy, at right angle to its disk, the radiant flux is observed. It appears such behavior is consistent with the proposed theoretical gravity model. Indeed, consider a massive galaxy with a dense nucleus in the center of it. (According to octonionic theory there are no black holes in nature, because in strong gravitational fields the D-bosons pairs are born and their gravitational description is unacceptable. Hence in case of a strong field the consideration could be restricted by flat Minkowski space, although assuming major vectors-potentials $A^c_\mu,c=4,5,6,7$ to be the main fields. From the representation proposed here $A^4$ and $A^7$ complete bound states, vacuum, and are therefore stable. So, restrict the consideration and research only $D$-bosons.)

For a particle falling ti the star interaction with $D$-boson is represented a gravitational field with Schwarzchild metrics (\ref{Shwm0}). Formally, denote particle mass by $m$, impulse by $p$ and energy by $p^0=E$ then there is the invariant
\begin{equation}\label{inve1}
g_{\mu\nu}p^{\mu\nu}=(1-\frac{r_g}r)E^2-\frac{p^2}{1-\frac{r_g}r}=m^2
\end{equation}
which means
\begin{equation}\label{inve10}
E^2=\frac{p^2}{(1-\frac{r_g}r)^2}+\frac{m^2}{1-\frac{r_g}r}
\end{equation}

It is evident from latter when reaching the gravitational radius its energy tends to infinity. Let $D$-boson mass be equal to Planck's mass. Then with $r\approx r_g+r_{pl}$ $D$-boson birth could happen in case the mass of the star piece enclosed into gravitational radius is estimated by $10^{12}$ sun mass.

Consider the massive vector field $A_\mu$
\begin{equation}\label{v1}
F^{\mu\nu}_{;\nu}-m^2A^\mu=0,\qquad F_{\mu\nu}=A_{\nu,\mu}-A_{\mu,\nu}.
\end{equation}

Rewrite using vector-potential $A_\mu$
\begin{equation}\label{v2}
\frac1{\sqrt{-g}}((A_{\sigma,\rho}-A_{\rho,\sigma})g^{\rho\mu} g^{\sigma\nu}\sqrt{-g}),\nu-m^2A^\mu=0.
\end{equation}

Introduce the scalar functions of radial coordinates and time $f^{lm}(r,t)$ and $h^{lm}(r,t)$, $k^{lm}(r,t), a^{lm}(r,t)$ \cite{WheelerRegger}
\begin{equation}\label{sph0}
A^0=f^{lm}(r,t)Y^{lm}(\theta,\phi),\qquad 
A^1=h^{lm}(r,t)Y^{lm}(\theta,\phi)
\end{equation}
\begin{equation}\label{sph2}
A^2=k^{lm}(r,t)Y^{lm}(\theta,\phi)+a^{lm}(r,t)Y^{lm}(\theta,\phi)/\sin\theta
\end{equation}
\begin{equation}\label{sph3}
A^3=k^{lm}(r,t)Y^{lm}(\theta,\phi)-a^{lm}(r,t)Y^{lm}(\theta,\phi)\sin\theta
\end{equation}

And for the radial component of $A_r$ it takes
\begin{equation}\label{urar1}
\lambda(h^{lm}-k^{lm}_{,r})+((f^{lm}_{,r}-h^{lm}_t)r^2)_{,t}f^{-1}
-m^2r^2h^{lm}=0
\end{equation}
where $\lambda=l(l+1)$. 

Assume the solution to be stable, then AЕ In this case solution exists with $A^0=A^2=A^3=0,A^1=A^1(t,\theta,\phi)$ and
\begin{equation}\label{urar2}
\omega^2(r)f^{-1}(r)+\frac\lambda{r^2}-m^2=0
\end{equation}

Hence, there is a solution $D=(0,C\cdot Y^{lm},0,0)$. That is centrally-symmetrical solution with a constant radius $r=r_g$. Which means the $D^{\pm}$-bosons pair birth happens. From the angular momentum conservation law they are born symmetrically relatively to the center of the star. Because of their nature, on the boundary of Schwarzchild radius the bosons form the boson condensate. Let the galaxy material be distributed like a rotating disk. If in the field of that system appears considerably large and slow galaxy and it is caught then the great thread of matter tends to the center of supermassive galaxy. As result is the increase in gravitational radius. The conversion of boson condensate from the level $\omega(r=r_{g_1})$ to the level $\omega(r=r_{g_2})$ happens. This conversion is accompanied with radiation because gravitational $D$-particles are charged. The angular distribution of radiation is estimated by the formula \cite{Davidov}
\begin{equation}\label{v9}
I=I_0\sum_{k=1}^la_k\cos^k\theta
\end{equation}

Once here $l=mpv\to\infty$ and the galaxy has a form of a disk then only the radiation orthogonally to the disk surface is possible. Hence, appears a radiation precisely at the right angle to the galaxy.

Along with discussed another scenario could take place. Let the supermassive galaxy be star-like. Vectorial $D$-bosons are born around the massive nucleus of the star, completing the bound state, which is close by its nature to the hydrogen atom. As a result the mass accumulation, as a boson cover around baryon nucleus, takes place. Due to the increase in the gravitational radius after the increase in mass the completion of more and more remote from center energy levels of the star happens. As a corollary of cooling process, higher levels of matter fall to the center. At the expense of increase in gravitational radius the radiation in all directions is emited, because no disk symmetry is assumed. As a result of it explosion and, perhaps, the birth of supernova happens.
\section{Conclusion}
In the physical review of the paper some phenomenological solutions of octonionic Lagrangian are proposed because the problem of physical interpretation of the octonionic Lagrangian deserves attention. The scope of the paper does not allow to research those solutions. Moreover, there are some more interesting octonionic Lagrangian physical corollaries. Particularly, it is amazing that vacuum state $u_0=\Psi_0$ and bound state $A^4_\mu A^{(7)\mu}$ are close by their structures and meanings. The author hopes the research of that fact would be useful in Higgs boson nature problem.

\end{document}